\documentstyle[12pt]{article}

\newcommand{\be}{\begin{equation}} 
\newcommand{\ee}{\end{equation}} 
\newcommand{\bref}[1]{(\ref{#1})} 
\newcommand{\ct}[1]{\cite{#1}}
\newcommand{\ctup}[1]{$^{\cite{#1}}$}

\def\theequation{\thesection.\arabic{equation}}
\def\@eqnnum{{\rm (\theequation)}} 
\def\secttit#1{\vglue 0.6cm{\bf\large\noindent{#1}}\vglue 0.2cm}

\def\NN{N\kern-.25em N} 
\def\rank{{\hbox{ \rm{rank} }}} 
\def\div{\, | \,} 
\def\lcm{\mathop{\rm lcm}} 
\def\ie{i.~e.} 
\def\fr#1#2{{{#1} \over {#2}}} 
\def\lrb#1{\left( { #1 } \right)} 
\def\lrcb#1{\left\{ { #1 } \right\} } 
\def\ex#1{e^{#1\pi i / N}} 
\def\Tr{{\rm{Tr}}} 
\def\gelement{ x_1^{n_1} x_2^{n_2} \dots x_d^{n_d} 
      \prod_{ij} z_{ij}^{n_{ij}} }
\def\gelementinv{ x_d^{-n_d} \dots x_2^{-n_2} x_1^{-n_1}
      \prod_{i<j} z_{ij}^{-n_{ij}} }

\def\sqr#1#2{{{\vbox{\hrule height#2pt 
        \hbox{\vrule width#2pt height#1pt \kern#1pt 
            \vrule width#2pt} 
        \hrule height#2pt}}}}

\begin{document}

\begin{titlepage}

%\noindent

%\begin{flushright}

MPI--PhT/94--88\\
CCNY--HEP--94/14\\
ITP--UH--21/94\\
hep-th/9412197\\
December 1994\\
Revised: October 1995\\

%\end{flushright} 

\vskip 1.6cm

\begin{center}

{\Large\bf The Solution of the }\\
\vskip0.5truecm 
{\Large\bf $d$-Dimensional Twisted Group Lattices}\\

\vskip 1.0cm

{ Olaf Lechtenfeld$^{*}$
\, and \, 
Stuart Samuel$^{\dagger}$}

\vskip 0.6cm

{\small 
$^*${\it Institut f\"ur Theoretische Physik, Universit\"at Hannover}\\
{\it Appelstra\ss{}e 2, 30167 Hannover, Germany}\\
{E-mail: lechtenf@itp.uni-hannover.de}\\
[0.4cm]

$^\dagger$
{\it Department of Physics}\\
{\it City College of New York}\\ 
{\it New York, NY 10031 USA}\\  
%{\it Max-Planck-Institut f\"ur Physik, Werner-Heisenberg-Institut}\\
%{\it F\"ohringer Ring 6, 80805 Munich, Germany}\\
{E-mail: samuel@scisun.sci.ccny.cuny.edu}\\
[0.5cm] 
} 
\normalsize
\vskip 1cm
\textwidth 6truein
{\bf Abstract}
\end{center}

\begin{quote}
\hspace{\parindent} 
{}\ \ \ The general $d$-dimensional twisted group lattice is solved.  
The irreducible representations of the corresponding group are constructed
by an explicit procedure. It is proven that they are complete. All matrix 
representation solutions to the quantum hyperplane equations are obtained.  

\end{quote} 

\vfill

%\textwidth 6.5truein
%\hrule width 5.cm

%{\small
%\noindent $^\#$
%Permanent address and address after August, 1995: \\ 
%\hspace*{0.4cm}Department of Physics, 
%City College of New York,\\ 
%\hspace*{0.4cm}New York, NY 10031, USA\\ 
%}    

\eject 
\end{titlepage} 

%\secttit{I.\ Introduction} 
{\bf\large\noindent{I.\ Introduction}}\vglue 0.2cm
\setcounter{section}{1}   
\setcounter{equation}{0}   
\label{s:i} 

The general $d$-dimensional twisted group lattice 
is a particular kind of group lattice.~\ctup{samuel90a,samuel91a}
A group lattice is a lattice 
constructed from a discrete group $G$ 
and a subset $\NN $ of $G$.  
The subset $\NN $ is called the nearest-neighbor set 
and is not necessarily a subgroup.  
Recall that a lattice 
is determined by specifying sites 
and bonds.  
For each element $g \in G$, 
one associates a site.  
Hence, the number of lattice sites is 
the number of elements of $G$.  
This is the order of $G$ and 
is denoted by $ o ( G ) $.  
The identity element $e$ of $G$ 
is by definition associated 
with the origin of the lattice.  
The subset $\NN$ 
is used to determine the bonds of the lattice.  
A site $g'$ is 
a nearest-neighbor site of $g$ 
if $g'g^{-1} \in \NN $.  
Hence,   
the nearest-neighbor sites of a site $g$ are 
the elements in the right-coset $\NN g$,    
so that the bonds ${\cal B}$ of the lattice are 
$ 
  {\cal B} =  
  \{ [ hg , g ] 
\ , \hbox{ such that } g \in G, h \in \NN  \} 
$.   
Here, the pair $[ hg , g ]$ indicates 
that a bond goes between sites $hg$ and $g$. 
One requires that 
if $h \in \NN $ then $h^{-1} \in \NN $ 
so that if $g'$ is a nearest neighbor of $g$ 
then  $g$ is also a nearest neighbor of $g'$.  
For more discussion on group lattices, 
see Refs.\ \ct{samuel90a}--\ct{samuel92a}.  

Group lattices are interesting 
for the following reason.  
If the irreducible representations 
of $G$ are known, 
then one can compute  
the free propagation of a particle 
through the lattice.  
The particle is allowed to move  
between nearest-neighbor sites.  
Hopping parameters are introduced 
to control the ease or difficulty 
of moving over a bond.  
The formalism is given 
in Ref.\ \ct{samuel90a} 
both for bosons or fermions.  
It is also possible to define field theories 
on group lattices.  
The free field theory is exactly solvable 
and a perturbative expansion exists  
for interacting theories.  
For details of the solution method, 
see Ref.\ \ct{samuel90a}.  

The group lattice is a general concept 
and is expected 
to have many physical applications.  
An example occurs for $C_{60}$. 
The carbon atoms sit at the sites of a group lattice 
based on the alternating group $A_5$.  
Using some approximations, 
a theoretical computation 
of the electronic structure of $C_{60}$
can be performed.  
There is good agreement between 
the group-lattice results 
and experiments.~\ctup{samuel93a,samuel93b}   

The {\it twisted\/} group lattices form 
a particular class of general group lattices.
To define the $d$-dimensional twisted group lattice, 
we must specify $G_d$ and $\NN$.  
The group $G_d$ has $d$ generators, 
denoted by $x_1, x_2, \dots , x_d$. 
They satisfy several relations.  
First, 
$x_i^{L_i} = e$,  
for $i = 1, 2, \dots, d$, 
where $L_i$ are positive integers.  
Secondly, 
it is useful to define 
\be 
 z_{ij}\ \equiv\ x_j^{-1} x_i^{-1} x_j x_i 
\ , \quad \quad 
{\hbox{ for }} i<j
\quad .   
\label{1.1}
\ee
The $ z_{ij} $ commute with the $x_k$, 
and consequently with themselves.  
In addition, 
$z_{ij}^{N_{ij}} = e$, 
where $N_{ij}$ are non-negative integers.  
Consistency requires that 
$N_{ij}$ and $N_{ji}$ divide  $L_i$ for $j \ne i$.  
Let us summarize these statements:  

Firstly, the group $G_d$ associated 
with the $d$-dimensional twisted group lattice is 
$$ 
  G_d\ =\ \Bigl\{ { x_1^{n_1} x_2^{n_2} \ldots x_d^{n_d}  }   
  \prod_{i<j} {z_{ij}^{n_{ij}}} 
    \ , {\hbox { such that }} n_i = 0, 1, \ldots  ,L_i{-}1 , 
$$
$$
   \qquad n_{ij}  =0,1,\ldots ,N_{ij}{-}1 , 
    \ x_j x_i x_j^{-1} x_i^{-1} = 
       z_{ij} \hbox{ for } i<j , 
    \ x_k z_{ij} = z_{ij} x_k , 
$$
\be
   \qquad\qquad {z_{ij}z_{kl} = z_{kl}z_{ij} , 
    \ x_i^{L_i} = 
    z_{ij}^{N_{ij}} = e , 
    \ N_{ij} \hbox{ divides } L_i \hbox{ and } L_j
   \hbox{ for all } i < j } \Bigr\} 
\quad .   
\label{1.2}
\ee

We remark that 
the group $G_d$ is an iterated semidirect product of cyclic groups.
First observe that $G_d$ contains an abelian normal subgroup,
$A_d\triangleleft G_d$, with
\be
   A_d\ =\ \Bigl\{ x_d^{n_d} \prod_{i=1}^{d-1} z_{id}^{n_{id}} \Bigr\}\
   =\ {\bf Z}_{L_d}\otimes{\bf Z}_{N_{1d}}
      \otimes\ldots\otimes{\bf Z}_{N_{d-1\ d}}
\quad.
\label{1.3}
\ee
{}From this it is clear that the quotient group $G_d{/}A_d$ 
is just the group associated to the twisted group lattice in
$d{-}1$ dimensions,
\be
   G_d/A_d\ =\ G_{d-1}\ \subset\ G_d
\quad.
\label{1.4}
\ee
One says that $G_d$ is an extension of $G_{d-1}$ by $A_d$.
Since $G_{d-1}$ is also a subgroup of $G_d$, one has 
$G_{d-1}\cap A_d=\{e\}$,
and $G_d$ is the semi-direct product of $G_{d-1}$ and $A_d$,
\be
   G_d\ =\ G_{d-1}\cdot A_d
\quad.
\label{1.5}
\ee
By iterating this construction,
one arrives at
\be
   G_d\ =\ \bigl( \ldots \bigl(\bigl( G_1\cdot A_2 
           \bigr) \cdot A_3 \bigr) \cdot \ldots \bigl) \cdot A_d
\quad,
\label{1.6}
\ee
with $G_1={\bf Z}_{L_1}$ and $A_i$ given by Eq.\ \bref{1.3}.

Secondly, the nearest-neighbor set $\NN$ 
associated with $G_d$
is chosen to be 
\be  
 \NN\ =\ \{ { 
   x_1, x_1^{-1}, x_2, x_2^{-1}, \dots , x_d, x_d^{-1} 
         } \} 
\quad .  
\label{1.7}
\ee 

Apparently,
a different group lattice is obtained for different 
values of $N_{ij}$ and $L_i$.  
The center of the group $G_d$
is generated by the $z_{ij}$.  
The number of elements in $G$ is 
\be  
   o (G_d)\ =\ 
   \biggl( \prod_{i=1}^{d} {L_i} \biggr) 
   \biggl( \prod_{i<j} {N_{ij}}  \biggr) 
\quad .  
\label{1.8}
\ee

The general form of an element is 
\be  
  g\ =\ \gelement 
\quad ,   
\label{a} 
\ee        
where $n_{i} $ ranges from $0$ to $L_{i} {-}1$ 
and   $n_{ij}$ ranges from $0$ to $N_{ij}{-}1$.  
Associate with a site $g$ 
of the form 
in Eq.\ \bref{a} 
the point 
\be  
 \lrb{ 
  n_1 , n_2, \ldots , n_d ;  
  n_{12}, n_{13},  \ldots , n_{d-1 \, d} 
     } 
\quad ,   
\label{1.10} 
\ee        
in the $d (d+1) / 2$ dimensional hypercubic lattice.  
We call the first $d$ coordinates 
the space coordinates 
and the last $d (d - 1) /2$ coordinates 
the internal coordinates.  

The nearest neighbors of $g$ 
are obtained by multiplying by $x_i$ or $x_i^{-1}$.  
When $g$ 
in Eq.\ \bref{a} 
is multiplyed  
by $x_i$,  
the exponent of $x_i$ 
increases by one 
and becomes $n_i{+}1$.% 
{\footnote{The exponents 
of the $z_{ij}$ 
may also change.}}   
This corresponds to a movement 
of one unit in the $i$th direction.  
When $g$ 
in Eq.\ \bref{a} 
is multiplyed  
by $x_i^{-1}$,  
the exponent of $x_i$ 
decreases by one 
and becomes $n_i{-}1$. 
This corresponds to a movement 
of one unit in the negative $i$th direction.  
If all $N_{ij} = 1$, 
then the regular periodic hypercubic lattice 
in $d$ dimensions is obtained,  
\ie\ $A_i={\bf Z}_{L_i}$ and 
$G_d=\bigotimes_{i=1}^d {\bf Z}_{L_i}$.
The twisted group lattices, 
for which $N_{ij} \ge 2$, 
differ from regular lattices 
in that one does not necessarily return 
to the same element when going around 
a closed path.  
For example, going around 
an elementary plaquette 
in the $i$-$j$ plane 
starting at the origin 
corresponds to 
$x_j^{-1} x_i^{-1} x_j x_i $, 
for $i<j$.  
This path does not return to the origin $e$ 
but to $z_{ij}$.  
In general, 
a path 
starting at the origin 
returns to the origin 
if and only if the region 
projected onto each of the $i$-$j$ planes 
has an area which is $0 \ ( \bmod \, {N_{ij}} )$.   
When particles propagate 
on a twisted group lattice, 
this constraint needs to be taken into account.   

The main goal of the current work 
is to solve the general $d$-dimensional group lattice.  
This requires that one obtains 
the irreducible representations of $G_d$ in Eq.\ \bref{1.2}.
In principle, all irreducible representations of $G_d$ can
be constructed from those of $G_{d-1}$ 
by using the method of induced representations.~\ctup{serre}
However, the explicit iteration of this method based on 
Eq.\ \bref{1.6} becomes rather involved.
In this paper, we have developed 
a different and more direct approach.
In Sect.\ 2, 
unitary representations are found. 
In Sect.\ 3, 
it is proven that 
the representations obtained 
in Sect.\ 2 
are irreducible and complete.  
With this solution, 
one can use the formulas 
in Refs.\ \ct{samuel90a,ls94a} 
to compute interesting results.  
The free propagation 
of particles on a $d$-dimensional twisted group lattice 
can be solved.  
If interacting field theories 
are put on these group lattices, 
a perturbative expansion  
can be performed.  

The cases of $d=2$, $d=3$ and $d=4$ 
have been solved 
in Refs.\ \ct{samuel91b,samuel92a,ls94a}.  
The solution method of the current work 
serves to reproduce 
the results for $d=2$, $d=3$ and $d=4$ 
in a unified framework.  
For general~$d$,
we uncover an interesting connection with 
alternating bilinear forms on free ${\bf Z}_N$--modules
of finite rank.
We develop this and the corresponding mathematics 
in Sect.\ 2.  
Because of the interesting physical character 
of the twisted group lattices 
there are likely to be applications 
in physics.  
One possibility worth exploring is generalized spin.  
The gamma matrices $\Gamma_i$ of the euclidean 
Dirac equation satisfy 
$\Gamma_i \Gamma_j + \Gamma_j \Gamma_i = \delta_{ij}$.  
This system of equations is closely related 
to the twisted group lattice 
with $ N_{ij} = 2$ for all $i < j$.  
Indeed, gamma matrix structure 
has been shown to arise naturally 
for this case.~\ctup{ls94a}   
Generalized gamma matrices for which 
$\Gamma_i \Gamma_j \pm \Gamma_j \Gamma_i = \delta_{ij}$ 
for various signs    
may be of interest.  
The results of 
Sect.\ 2 
allow one to construct 
the irreducible representations 
for this algebra.  
Equations such as 
$ 
   \Gamma_i \Gamma_j + 
  \exp \lrb{i \phi_{ij}}  
   \Gamma_j \Gamma_i = 0 
$, 
where $\exp \lrb{i \phi_{ij}} $ are roots of unity,  
are of interest in parastatistics.  
Again, 
the results 
of Sect.\ 2 
may be of use.  

At an intermediate step in our solution method, 
we are lead to the equation 
\be  
  \Gamma_i\ \Gamma_j\ =\ 
  \Gamma_j\ \Gamma_i\ 
    \exp \lrb{2 \pi i n_{i j} / N } 
\quad ,  
\label{ab} 
\ee  
where $N$ and $n_{i j} = - n_{j i}$ are integers, 
and where   
$\Gamma_i$ are matrices.  
The equations 
in Eq.\ \bref{ab} 
enter in several mathematical and physical contexts.  
When $\Gamma_i$ are regarded 
as elements of an abstract algebra, 
\bref{ab} 
corresponds to the quantum hyperplane, 
which arises in regard  
to quantum groups.~\ctup{manin88a,wz90a,wzos90a,filk} 
Hence, matrix solutions to 
\bref{ab} 
are matrix representations of the quantum hyperplane algebra. 
When the $\Gamma_i$ are $N \times N$ matrices, 
where $N$ is the parameter on the right-hand side of 
Eq.\ \bref{ab},  
there are several applications: 
(i) the solution to self-dual Yang-Mills equations 
on a hypertorus 
with twisted boundary conditions \ct{thooft79},
(ii) Witten's work \ct{witten82} 
on contraints on supersymmetry breaking, 
(iii) the twisted Eguchi-Kawai model 
\ct{go83} 
and 
(iv) the twisted Eguchi-Kawai model at finite temperature 
\ct{kb84,fa84}. 
Eq.\ \bref{ab} is also relevant for 
certain $d$-dimensional crystals 
with screw dislocations.~\ctup{samuel92a} 

Problem (i)
was considered by G.\ 't Hooft
in connection with quark confinement
in an $SU(N)$ gauge theory.
In Ref.\ \ct{thooft81},
solutions were found
in $d=4$
for the special case
in which $N$ is not divisible
by a prime-number squared.
Complete solutions in arbitrary dimensions
were found in refs.\ \ct{bg86,lp86}. 
Our results for the intermediate problem governed by 
Eq.\ \bref{ab} 
are equivalent to those in refs.\ \ct{bg86,lp86}. 

The solution to Eq.\ \bref{ab} 
for $\Gamma_i$ of arbitrary size 
makes use of  
$N \times N$ matrices 
$P_{\left( N \right)}$ and $Q_{\left( N \right)}$   
satisfying  
\be  
  P_{\left( N \right)}\ Q_{\left( N \right)}\ =\  
  Q_{\left( N \right)}\ P_{\left( N \right)}\ 
    \exp \left( {{{2\pi i} \over N}} \right) 
\quad .  
\label{b} 
\ee  
An explicit realization is\ctup{thooft78} 
$$ 
  P_{(N)}\ =\    
    \left( \matrix{ 
   0  &1  &0  &0  &\ldots   &0  &0  \cr
   0  &0  &1  &0  &\ldots   &0  &0  \cr
   0  &0  &0  &1  &\ldots   &0  &0  \cr
   \vdots  &\vdots  &\vdots  &\vdots  &\ddots   &\vdots  &\vdots  \cr
   0  &0  &0  &0  &\ldots   &0  &1  \cr
   1  &0  &0  &0  &\ldots   &0  &0  \cr
    } \right)  
\quad , 
$$ 
\be  
  Q_{(N)}\ =\    
    \left( \matrix{ 
   1  &0  &0  &0  &\dots   &0  \cr
   0  &\ex{2}  &0  &0  &\ldots   &0  \cr
   0  &0  &\ex{4}  &0  &\ldots   &0  \cr
   \vdots  &\vdots  &\vdots  &\vdots  &\ddots   &\vdots  \cr
   0  &0  &0  &0  &\ldots   &\ex{2(N-1)}  \cr
    } \right)  
\quad ,  
\label{c} 
\ee  
where the subscript $\lrb{N}$ indicates 
the size of the  matrix.  
The $N \times N$ identity matrix is denoted by $I_{\lrb{N}}$.  

Because of the need for elaborate notation, 
we summarize some additional conventions here.  
When curly brackets $\{ \ \}$ appear 
around a variable with $ij$ indices, 
it means that $ij$ range over 
the $d(d{-}1)/2$ values 
$ ij=12, 13, \dots , d{-}1\,d$.   
Integers $k_i$ and $k_{ij}$ are used 
for representation labels.  
The letter $p$ indicates a prime number.  
The integer $\ell$ labels different prime sectors.  
These and other superscripts are enclosed 
in parenthesis to avoid possibly confusion 
with an exponent.  
The letters $r$, $s$, $t$ and $u$, with or without 
subscripts and superscripts, are reserved for integer powers.  
The reader should be aware that 
the use of the exponents $r$ and $u$ differs 
from Ref.\ \ct{ls94a}. 
Representations are denoted by $(r)$ and $(s)$ 
and are enclosed in parenthesis to distinguish 
them from the exponents $r$ and $s$.  
For the corresponding dimension and character, 
the letters $d$ and $\chi$ are used.   
A capital greek $\Omega$ 
indicates a two-form.  
The associated antisymmetric matrix 
is denoted by $\omega$.  
The standard one-form basis is 
$dx^1, dx^2, \dots , dx^d$.   
The greek letters 
$\alpha$ and $\beta$ stand for general one-forms.  
The corresponding component, 
\ie, the coefficient of $dx^i$,  
is denoted by using the equivalent latin index 
$a_i$ and $b_i$.  
With a subscript $i$ or $ij$,  
$n$ stands 
for a power of a group element factor  
as in Eq.\ \bref{a}.  
Without a subscript, 
it labels different tensor product sectors.  
In the $\ell$th prime sector, 
$n$ ranges from $1$ to $N_\ell$.  
One should not confuse $N_\ell$ 
with the twisting integers $N_{ij}$ 
which determine the group $G_d$ 
in Eq.\ \bref{1.2}.  
Modular arithematic is indicated 
with the symbol ``$\bmod$''.  
Because the construction procedure is algebraically tedious, 
we have provided an example of the method 
in an appendix.

\secttit {II. Irreducible Representation Construction} 
\label{s:irc} 
\setcounter{section}{2}   
\setcounter{equation}{0}   

%\hspace{\parindent} 
This section finds  
the complete set of irreducible representations 
for the $d$-dimensional twisted group lattice  
defined by the set $\{N_{ij},L_i\}$.
The representations are determined by the two sets of integers 
$k_{ij}$, for $1 \le i < j \le d$,  
and $k_{i}$, for $i=1, \ldots , d$.  
Hence, $d (d+1)/2$ integers  
specify a representation $\lrb{r}$.  
The $k_{ij}$ are ``internal momenta'' 
associated with the $z_{ij}$ group elements 
and range from $0$ to $N_{ij}{-}1$.  
The $k_{i}$ are space-time momenta 
and take on values from $0$ to $L_i{-}1$. 
However, some sets of $k_{i}$ lead to equivalent representations.  
Equivalent sets are related by a group $E$ 
which is specified below.  
It turns out that the dimension $d_{\lrb{r}}$ of $\lrb{r}$ 
depends only on the ratios $k_{ij}/N_{ij}$ 
and not on the $k_i$.   

Since the $z_{ij}$ commute with all elements, 
the corresponding representation matrices 
can be chosen to be diagonal 
\be  
  z_{ij}\ \rightarrow\ I_{ \lrb{d_{\lrb{r}}} } 
   \exp \left( { 2 \pi i { {k_{ij} } \over { N_{ij} } } } \right) 
\quad .  
\label{2.1} 
\ee  
The phases 
in Eq.\ \bref{2.1}  
are fixed by the requirement that 
$z_{ij}^{ N_{ij}}$ be the identity element.  
The matrix representations of $x_j$ 
are of the form\ctup{samuel92a}
\be  
  x_j\ \rightarrow\ \Gamma_j 
   \exp \left( { 2 \pi i { {k_{j} } \over { L_{j} } } } \right) 
\quad , 
\label{2.2} 
\ee  
where the $\Gamma_i$ are 
$d_{\lrb{r}} \times d_{\lrb{r}}$ matrices 
satisfying 
\be  
  \Gamma_j \Gamma_i\ =\ \Gamma_i \Gamma_j \ 
    \exp \left( {2\pi i{{k_{ij}} \over {N_{ij}}}} \right) 
\ , \quad \quad 
{\hbox {for}} \ 1 \le i < j \le d  
\quad , 
\label{2.3} 
\ee  
and 
\be  
  \Gamma_i^{L_i} = 1 
\ , \quad \quad 
{\hbox {for}} \ i = 1, 2, \ldots , d  
\quad . 
\label{2.4} 
\ee  
The key problem is to find $\Gamma_i$ satifying 
Eqs.\ \bref{2.3} and \bref{2.4}.  
Note that the $k_i$ no longer appear.

For the rest of this section, 
fix the values of the $\{ k_{ij} \}$.  
Since it is only the ratio 
$  {{k_{ij}}  / {N_{ij}}} $ 
which enters 
Eqs.\ \bref{2.1} and \bref{2.3}, 
we reduce this fraction 
by removing common factors:  
\be  
  {{k_{ij}}  / {N_{ij}}} = 
  {{k_{ij}'} / {N_{ij}'}} 
\quad , 
\label{2.5} 
\ee 
where 
\be  
  \gcd \left( {k_{ij}',N_{ij}'} \right) = 1  
\quad . 
\label{2.6} 
\ee  
When $k_{ij} = 0$, 
we set $k_{ij}' = 0$ and define $N_{ij}' = 1$. 
Let $N'$ be the least common multiple of the $N_{ij}'$, 
\be  
  N' \equiv \lcm \left( { \{ { N_{ij}' } \} } \right) 
\quad . 
\label{2.7} 
\ee  

At this point 
we use the prime factorization idea  
of Ref.\ \ct{ls94a}.  
Factorize $N'$ and the $N_{ij}'$ into primes $p_\ell$ using    
\be  
  N' = \prod_{\ell=1}^L {p_{\ell}^{s_{\ell}}} 
\ , \quad \quad 
{\hbox {where}} \ s_{\ell} \ge 1   
\quad , 
\label{2.8} 
\ee  
and 
\be 
  N_{ij}' = 
     \prod_{\ell=1}^L 
        {p_{\ell}^{t_{ij}^{\left( {\ell} \right)}}} 
\ , \quad \quad 
{\hbox {where}} \ t_{ij}^{\left( {\ell} \right)} \ge 0    
\quad . 
\label{2.9} 
\ee  
The $L$ prime numbers 
which enter in $N'$ are 
$p_1$, $p_2$, $\ldots$, $p_L$. 
The integer     
$ 
  t_{ij}^{\left( {\ell} \right)} 
$ 
is defined to be zero  
when $p_{\ell}$ 
does not appear in $N_{ij}'$.  
Eq.\ \bref{2.7} implies that 
$s_{\ell}$ is 
the maximum value 
of the $t_{ij}^{\left( {\ell} \right)}$ 
as $ij$ ranges over $1 \le i < j \le d$, \ie,  
\be  
  s_{\ell} = \mathop{\max }\limits_{\left\{ {ij} \right\}} 
   \left\{ {t_{ij}^{\left( {\ell} \right)}} \right\}    
\quad . 
\label{2.10} 
\ee  
The ratio 
in Eq.\ \bref{2.5} 
can be written as a sum of ``prime fractions''  
via 
\be  
   {{k_{ij}'} \over {N_{ij}'}}\ =\ 
   \sum\limits_{\ell=1}^L 
  {{{\omega_{ij}^{\left( {\ell} \right)}} \over {p_{\ell}^{s_{\ell}}}}}  
\quad . 
\label{2.11} 
\ee  
Eq.\ \bref{2.11} 
defines the integers 
$ 
  \omega_{ij}^{\left( {\ell} \right)} 
$.   
It may happen that 
the fraction 
$\omega_{ij}^{\left( {\ell} \right)} / p_{\ell}^{s_{\ell}}$ 
can be reduced,  
in which case  
$ p \div \omega_{ij}^{\left( {\ell} \right)} $,  
but if
$
  t_{ij}^{\left( {\ell} \right)} =  s_{\ell} 
$ 
for a particular value of $ij$,  
then, as consequence of 
Eqs.\ \bref{2.8}--\bref{2.10},  
$ 
  \gcd \left( { 
             \omega_{ij}^{\left( {\ell} \right)} , 
             p_{\ell}
               }  \right) = 1 
$. 
In short, 
there is at least one 
$\omega_{ij}^{\left( {\ell} \right)}$ 
that is relatively prime to $p_{\ell}$.

The results in the previous paragraph
allow one to factorize the $\Gamma_i$.   
Write $\Gamma_i$ as a tensor product of $L$ factors via 
\be  
  \Gamma_i \ = \           \Gamma_i^{\left( 1 \right)} 
                   \otimes \Gamma_i^{\left( 2 \right)} 
    \otimes \ldots \otimes \Gamma_i^{\left( L \right)} 
\quad .  
\label{2.12} 
\ee  
If 
\be  
  \Gamma_j^{\left( {\ell} \right)} \
  \Gamma_i^{\left( {\ell} \right)} \,=\, 
  \Gamma_i^{\left( {\ell} \right)} \
  \Gamma_j^{\left( {\ell} \right)} \ 
  \exp \left( {2\pi i 
  {{\omega_{ij}^{\left( {\ell} \right)}} \over {p_{\ell}^{s_{\ell}}}}} 
       \right) 
\label{2.13} 
\ee  
and 
\be  
  \left( {\Gamma_i^{\left( {\ell} \right)}} \right)^{L_i} = 1  
\quad , 
\label{2.14} 
\ee   
then Eqs.\ \bref{2.3} and \bref{2.4} are satisfied 
as a consequence of Eqs.\ \bref{2.11}--\bref{2.14}, 
as one can easily check.  
The converse is also true.
In summary, 
the problem has been reduced 
to finding representations 
of Eqs.\ \bref{2.13} and \bref{2.14}, \ie\ 
of a single prime-factor sector $\ell$.  

Since it is sufficient to consider a single sector, 
let us temporarily simplify notation 
by doing away with the index $\ell$.  
Define  
\be  
  \omega_{ij} \equiv \omega_{ij}^{\left( {\ell} \right)} 
\ , \quad \quad 
             s\equiv s_{\ell} 
\ , \quad \quad  
             p\equiv p_{\ell} 
\quad , 
\label{2.15} 
\ee   
so that 
\be  
  \Gamma_j^{\left( {\ell} \right)} \
  \Gamma_i^{\left( {\ell} \right)} \ =\ 
  \Gamma_i^{\left( {\ell} \right)} \
  \Gamma_j^{\left( {\ell} \right)} \ 
  \exp \left( {2\pi i{{\omega_{ij}} \over {p^s}}} \right) 
\ , \quad \quad   
1 \le i < j \le d 
\quad . 
\label{2.16} 
\ee   
The $\omega_{ij}$ are integers modulo $p^s$.  
One has 
\be  
  \gcd \left( {\left\{ {\omega_{ij}} \right\},p} \right) = 1 
\qquad 
{\hbox{ for at least one value of }}ij 
\quad . 
\label{2.17} 
\ee  
If is convenient to define 
\be  
  \omega_{ji} \equiv - \omega_{ij}  \pmod {p} 
\ , \quad \quad   \hbox{ for }i<j 
\quad .  
\label{2.18} 
\ee   
Then, the integers $\omega_{ij}$ comprise 
an antisymmetric $d \times d$ matrix, $\omega$.  
Such a matrix can be associated with 
a formal two-form $\Omega$ defined by 
\be  
  \Omega\ \equiv  
    \sum\limits_{1 \le i < j \le d} 
    {\omega_{ij}}\ dx^i \wedge dx^j 
\quad .  
\label{2.19} 
\ee   
{}From Eq.\ \bref{2.17}, 
one concludes $\Omega \ne 0 \lrb{ \bmod \ {p} } $. 

We now begin to decompose the two-form~$\Omega$.
More precisely, we shall show that
there exist $2M$ independent one-forms 
$
  \alpha^{\left( 1 \right)} 
$, 
$ 
   \beta^{\left( 1 \right)} 
$, 
$
  \alpha^{\left( 2 \right)} 
$, 
$  
   \beta^{\left( 2 \right)} 
$, 
$ 
  \ldots  
$, 
$ 
  \alpha^{\left( M \right)} 
$, 
$  
   \beta^{\left( M \right)} 
$,  
such that 
\be  
 \Omega\ =\ 
  \alpha^{\left( 1 \right)} \wedge \beta^{\left( 1 \right)} + 
  \alpha^{\left( 2 \right)} \wedge \beta^{\left( 2 \right)} + 
   \ldots + 
  \alpha^{\left( M \right)} \wedge \beta^{\left( M \right)} + 
  \Omega^{\prime} \pmod {p^s}  
\quad .   
\label{2.20} 
\ee   
Here, $M$ is the smallest integer such that
\be  
 \Omega^{{\wedge}(M+1)}\ \equiv\ 
 \underbrace{
  \Omega\wedge\Omega\wedge\ldots\wedge\Omega
   }_{M{+}1 {\rm\ times}}
 \ =\ 0 \pmod {p}
\quad ,
\label{2.21} 
\ee   
and 
\be  
\Omega^{\prime}\ =\ 0 \pmod {p} 
\quad .  
\label{2.22} 
\ee    

We note that
$2M$ is equal to the rank of the matrix 
$\omega \left( { \bmod \ {p} } \right)$.
It makes sense to consider the rank of a matrix modulo $p$ 
because the integers modulo $p$ constitute a field.  
It is also a well-known result that 
the rank of an antisymmetric matrix 
is an even integer.   
If we write 
\be  
  \Omega^{\prime} \ 
  \equiv \sum\limits_{1 \le i < j \le d} 
  {\omega^{\prime}_{ij}}\ dx^i \wedge dx^j 
\quad ,   
\label{2.23} 
\ee   
then Eq.\ \bref{2.22} implies that
\be  
  p \div \omega^{\prime}_{ij} 
\ , \quad \quad   \hbox{ for all }i<j 
\quad .    
\label{2.24} 
\ee   
We express  
Eq.\ \bref{2.24} 
compactly as 
$ 
  p \div \Omega^{\prime} 
$.  
When we say that one-forms 
$
  \omega^{\left( 1 \right)} 
$,  
$  
  \omega^{\left( 2 \right)} 
$,  
$  
  \ldots  
$,  
$ 
  \omega^{\left( N \right)} 
$   
are independent, 
we mean that 
the only solution to the equation 
$ 
 \sum_{n=1}^N c_n \omega^{\left( n \right)} = 
  0 \left( { \bmod \ {p} } \right) 
$ 
is 
$c_n = 0 \left( { \bmod \ {p} } \right) $.% 
{\footnote 
{Equivalently, 
$
  \omega^{\left( 1 \right)},  
  \omega^{\left( 2 \right)},  
  \ldots , 
  \omega^{\left( N \right)} 
$ 
are independent if and only if   
$ 
  \omega^{\left( 1 \right)} \wedge 
  \omega^{\left( 2 \right)} \wedge 
                     \ldots \wedge 
  \omega^{\left( N \right)}  
  \ne  0 \left( { \bmod \ {p} } \right)  
$.}}    

For $s=1$, 
Eq.\ \bref{2.20} reduces to the standard decomposition of a two-form
over a number field (in this case ${\bf Z}_p$).
When $s>1$, it is non-standard because the abelian ring ${\bf Z}_{p^s}$
contains zero divisors, namely $p^u$ for $u=1,\dots,s{-}1$.
By hiding the zero divisor part of $\Omega$ in $\Omega'$, 
we can apply the standard decomposition to $\Omega{-}\Omega'$.

A constructive proof of Eqs.\ \bref{2.20}--\bref{2.22} is as follows.
If $M=0$, let $\Omega^{\prime} = \Omega$. 
Then, there is nothing to show:
If $M=0$ then $\Omega = 0\left( { \bmod \ {p} } \right) $ 
so that $\Omega^{\prime} = 0 \left( { \bmod \ {p} } \right) $. 
Thus, assume $M \ge 1$.  
Since $\rank \omega \left( { \bmod \ {p} } \right) = 2M$, 
there exists a least one entry $\omega_{ij}$ 
such that $\omega_{ij} \ne 0 \left( { \bmod \ {p} } \right) $.  
By relabelling indices we may assume that ${ij} = 12$ 
without loss of generality.  
Let 
$$ 
  a_1^{\left( 1 \right)} \equiv 1 
\ , \quad \quad  
  a_2^{\left( 1 \right)} = 
  b_1^{\left( 1 \right)} \equiv 0 
\quad , 
$$ 
$$ 
  a_i^{\left( 1 \right)} \equiv 
   - \omega_{2i} / \omega_{12} \pmod {p^s} 
\ , \quad \quad  i\ge 3 
\quad , 
$$ 
\be  
  b_i^{\left( 1 \right)} \equiv 
  \omega_{1i} 
\ , \quad \quad  i\ge 2 
\quad .    
\label{2.25} 
\ee   
In defining $a_i^{\left( 1 \right)}$, 
it makes sense to divide by $\omega_{12}$, 
because $\omega_{12}^{-1}$ exists 
since $\gcd \left( { \omega_{12} , p } \right) = 1 $.  
Take 
\be  
  \alpha^{\left( 1 \right)}\ \equiv\ 
  \sum\limits_{i=1}^d a_i^{\left( 1 \right)} dx^i 
\ , \quad \quad  
  \beta^{\left( 1 \right)}\ \equiv\ 
  \sum\limits_{i=1}^d b_i^{\left( 1 \right)} dx^i 
\quad .    
\label{2.26} 
\ee   
Note that 
$ 
  \gcd \left( { b_2^{\left( 1 \right)} , p } \right) = 
  \gcd \left( {     \omega_{12}        , p } \right) = 1 
$.  
Define the remainder two-form $\hat \Omega$ by 
\be  
  \hat \Omega\ =\ 
  \Omega - 
  \alpha^{\left( 1 \right)} \wedge \beta^{\left( 1 \right)} 
\quad .    
\label{2.27} 
\ee   
A short computation shows that 
\be 
  \hat \Omega\ = 
  \sum\limits_{3\le i<j\le d} \hat \omega_{ij}\ dx^i \wedge dx^j 
\quad ,     
\label{2.28} 
\ee   
so that 
$\hat \Omega$ involves only the differentials 
$ 
 dx^{3},dx^{4}, \ldots ,dx^d 
$.   
Repeat the previous construction for $\hat \Omega$ 
and continue until the process terminates.  
The process must terminate 
because the rank of $\omega$ is finite. 
Eventually, 
some remainder matrix has zero rank modulo $p$.  
One then obtains the desired result 
in Eq.\ \bref{2.20}. 
After a relabelling of the index $i$ for $dx^i$, 
one sees that 
$ 
 \alpha^{\left( m \right)}
$ 
involves the differentials 
$ 
dx^{2m-1}, dx^{2m+1}, dx^{2m+2}, \ldots , dx^d 
$,  
and that 
$ 
  \beta^{\left( m \right)}
$ 
involves the differentials 
$ 
 dx^{2m} ,dx^{2m+1}, dx^{2m+2}, \ldots , dx^d 
$.  
Hence, 
the forms 
$  
\alpha^{\left( 1 \right)} 
$, 
$ 
\beta^{\left( 1 \right)} 
$, 
$   
\alpha^{\left( 2 \right)} 
$, 
$   
\beta^{\left( 2 \right)} 
$, 
$ 
\ldots 
$, 
$  
\alpha^{\left( M \right)} 
$ 
and 
$  
\beta^{\left( M \right)} 
$ 
are independent modulo $p$. 
As a consequence, their total wedge product is non-zero modulo $p$,
proving that, modulo $p$, 
$\Omega^{{\wedge}M}\ne 0$ but
$\Omega^{{\wedge}(M+1)}=0$.
This completes the proof. 

An examination of this proof 
reveals that it is both inductive and constructive.  
Take    
$ 
  a_i^{\left( m \right)},   
  b_i^{\left( m \right)},  
  m = 1, \ldots , M 
$,  
to be  
\be  
  \alpha^{\left( m \right)}\ \equiv\  
   \sum\limits_{i=1}^d a_i^{\left( m \right)} dx^i 
\ , \quad \quad 
   \beta^{\left( m \right)}\ \equiv\ 
   \sum\limits_{i=1}^d b_i^{\left( m \right)} dx^i 
\quad .      
\label{2.29} 
\ee   
Define $d$-component vectors 
$\vec a^{\left( m \right)}$ and 
$\vec b^{\left( m \right)}$ 
by 
$$ 
  \vec a^{\left( m \right)}\ \equiv\ 
 \left( { a_1^{\left( m \right)} , 
          a_2^{\left( m \right)} , \ldots , 
          a_d^{\left( m \right)}  
 } \right) 
\ , \quad \quad 
  \vec b^{\left( m \right)}\ \equiv\ 
 \left( { b_1^{\left( m \right)} , 
          b_2^{\left( m \right)} , \ldots , 
          b_d^{\left( m \right)}  
 } \right) 
\quad . 
$$    
Then, the $2M$ vectors 
$ 
  \vec a^{\left( m \right)},   
  \vec b^{\left( m \right)} 
$ 
are independent modulo $p$.  
The rank of $\omega$ modulo $p$ 
is the number of 
$\vec a^{\left( m \right)}$ and 
$\vec b^{\left( m \right)}$ vectors, 
as one can verify.  
Hence, this is indeed $2 M$ modulo $p$.  
We call the process of obtaining 
$  
  \alpha^{\left( 1 \right)},  
   \beta^{\left( 1 \right)},  
  \alpha^{\left( 2 \right)},  
   \beta^{\left( 2 \right)},  \ldots , 
  \alpha^{\left( M \right)}, 
   \beta^{\left( M \right)} 
$ 
the {\it Sympletic Construction} for $\Omega$.  

The decomposition of $\Omega$, as in Eq.\ \bref{2.20},
is not yet complete since, modulo ${\bf Z}_{p^s}$,
the remainder $\Omega^{\prime}$ is nonzero.
However, we may continue after dividing $\Omega^{\prime}$ by~$p$,
which must be possible according to Eq.\ \bref{2.22}.
Repeated applications of Eqs.\ \bref{2.20}--\bref{2.22} leads 
to the result that
any two-form $\Omega$ (as in Eq.\ \bref{2.19}) can be decomposed as
\be 
 \Omega\ =\ 
  \sum\limits_{n=1}^N {}p^{u_n}\ 
   \alpha^{\left( n \right)} \wedge 
    \beta^{\left( n \right)} \pmod {p^s} 
\quad ,     
\label{2.30} 
\ee   
where 
$ 
  0 \le u_n \le s-1 
$,  
$ 
  u_1 = 0 
$, 
and 
$\alpha^{\left( n \right)},\beta^{\left( n \right)}$ 
are independent one-forms. 

It is noteworthy that the $d\times d$ matrix of row vectors
$\vec a^{\left(1\right)}, \vec b^{\left(1\right)}, \ldots,
\vec a^{\left(N\right)}, \vec b^{\left(N\right)}$
(filled up with additional null vectors if $2N<d$)
constitutes essentially the transformation matrix~$X$
employed in refs.\ \ct{bg86,lp86}.

For a proof,
apply the sympletic construction to $\Omega$ to obtain 
\be  
  \Omega \equiv \Omega_0\ =\ 
   \alpha^{\left( 1 \right)} \wedge 
    \beta^{\left( 1 \right)} + 
   \alpha^{\left( 2 \right)} \wedge 
    \beta^{\left( 2 \right)} + \ldots + 
   \alpha^{\left( {M_0} \right)} \wedge 
    \beta^{\left( {M_0} \right)} + 
   \Omega^{\prime}_0 \pmod {p^s}  
\quad . 
\label{2.31} 
\ee   
Put 
$$ 
  u_1 = u_2 = \ldots = u_{M_0} = 0 
\quad . 
$$ 
Note that 
$ 
  M_0 \ge 1  
$ 
because 
$ 
\Omega_0 \ne 0 \left( { \bmod \ {p} } \right)  
$.  
Let  
$$ 
  \Omega_1 \ =\ 
  \Omega^{\prime}_0/p 
  \pmod {p^{s-1}}  
\quad . 
$$ 
It makes sense to divide by $p$ because 
$p \div \Omega^{\prime}_0$ 
since $\Omega^{\prime}_0 = 0 \left( { \bmod \ {p} } \right) $, \ie\ 
every element of the matrix $\omega^{\prime}_0$ 
is divisible by $p$.  
Apply the sympletic construction to 
$\Omega_1$:
$$ 
  \Omega_1\ =\
  \alpha^{\left( {M_0+1} \right)} \wedge 
   \beta^{\left( {M_0+1} \right)} + \ldots + 
  \alpha^{\left( {M_0+M_1} \right)} \wedge 
   \beta^{\left( {M_0+M_1} \right)} + 
  \Omega^{\prime}_1 
    \pmod {p^{s-1}}  
\quad . 
$$ 
It may be that $M_1 = 0$ because 
$\Omega_1 = 0 \left( { \bmod \ {p} } \right) $ 
already; 
then 
$ 
 \Omega_1 = \Omega^{\prime}_1 
$.    
If $M_1 > 0$, 
put 
$$ 
  u_{M_0+1} = u_{M_0+2} = \ldots = u_{M_0+M_1} = 1 
\quad . 
$$ 
Since 
$ 
 \Omega^{\prime}_1 = 
  0 \left( { \bmod \ {p} } \right) 
$, 
one can define  
$$ 
  \Omega_2\ =\
  \Omega^{\prime}_1/p 
   \pmod {p^{s-2}} 
$$ 
and apply the sympletic construction to 
$\Omega_2$,  
and so on.  
At each stage, $M_u$ counts the number of
$u_n$-values equal to $u$.
The process must terminate 
with $\Omega'_{s-1}=0$ and $\sum_{u=0}^{s-1} M_u \equiv N$, 
because the maximum number 
of independent one-forms is $d$.  
Hence, $N \le {d \over 2}$ for $d$ even 
and $N \le { {d-1} \over 2}$ for $d$ odd. 
By these means, 
sympletic constructions for 
$\Omega_u$, $u=0,\ldots,s{-}1$,
are obtained,
with $2M_u=\rank\omega_u\ (\bmod\ p)$. 
One reconstructs $\Omega^{\prime}_u$   
from $\Omega_{u+1}$ 
by using 
$ 
  \Omega'_u = 
    p \Omega_{u+1} 
  \left( { \bmod \ p^{s-u} } \right)
$ 
for $u=s{-}1,\ldots,0$,
beginning with
$\Omega^{\prime}_{s-1}=0$.
Then, one recursively expresses these two-forms as summands 
$\alpha^{\left( n \right)} \wedge \beta^{\left( n \right)}$   
to arrive at 
Eq.\ \bref{2.30}. 
The proof is completed.  

Let us now restore the index $\ell$.  
For each $\ell$ there is a two-form $\Omega^{\lrb{\ell}}$ 
of the form 
\be 
  \Omega^{\lrb{\ell}}\ =\ 
   \sum\limits_{n=1}^{N_{\ell}} p_{\ell}^{u_{\ell n}}\ 
   \alpha^{\lrb{\ell n}} \wedge 
    \beta^{\lrb{\ell n}} \pmod {p_{\ell}^{s_{\ell}}} 
\quad .    
\label{2.32} 
\ee   
We define 
$ a_i^{\lrb{\ell n}}$ and 
$ b_i^{\lrb{\ell n}}$ via  
\be  
  \alpha^{\lrb{\ell n}}\ \equiv\ 
   \sum\limits_{i=1}^d 
     a_i^{\lrb{\ell n}} dx^i 
\ , \quad \quad 
   \beta^{\lrb{\ell n}}\ \equiv\ 
  \sum\limits_{i=1}^d 
     b_i^{\lrb{\ell n}} dx^i 
\quad .   
\label{2.33} 
\ee   
Actually, 
for a fixed $\ell$, 
by relabelling the index $i$,  
one may assume, without loss of generality, that 
the sum in $i$ for 
$\alpha^{\lrb{\ell n}}$ 
starts at $2n{-}1$ 
and 
the sum in $i$ for 
$ \beta^{\lrb{\ell n}}$ 
starts at $2 n$ 
and that 
\be  
      a_{2n-1}^{\lrb{\ell n}} = 1 
\ , \quad \quad  
      a_{2n  }^{\lrb{\ell n}} = 0 
\ , \quad \quad  
  \gcd 
 \lrb{b_{2n  }^{\lrb{\ell n}},p} \ne 0 
\quad . 
\label{2.34} 
\ee   
These results follow from the details of the symplectic construction 
and imply that, 
within the $\ell$th sector, 
the vectors 
$ 
  \vec a^{\lrb{\ell n}}, 
  \vec b^{\lrb{\ell n}}, 
$ 
for 
$ 
n = 1, \ldots , N_{\ell},  
$ 
are independent modulo $p$.  

We now claim that
\be  
  \Gamma_i^{\left( {\ell} \right)}\ =\ \
  \bigotimes_{n=1}^{N_{\ell}} \Biggl( 
  Q_{ \lrb{p_{\ell}^{ 
         r_{\ell n}}} }^{a_i^{\lrb{\ell n}}}\ 
  P_{ \lrb{p_{\ell}^{ 
         r_{\ell n}}} }^{b_i^{\lrb{\ell n}}}\ 
  \exp \biggl[ {-i\pi \left( {L_i{-}1} \right) 
  {{a_i^{\lrb{\ell n}}
    b_i^{\lrb{\ell n}}} \over 
    {p_{\ell}^{r_{\ell n}}}}} \biggr] 
  \Biggr)
\label{2.35} 
\ee   
is a solution to Eqs.\ \bref{2.13} and \bref{2.14}.
Here, $\bigotimes_{n=1}^{N_{\ell}}$ indicates 
a tensor product of $N_{\ell}$ matrices, 
and the powers 
$a_i^{\lrb{\ell n}}$ 
and 
$b_i^{\lrb{\ell n}}$ 
of the matrices 
$ Q_{ \lrb{p_{\ell}^{r_{\ell n}}} }$ 
and 
$ P_{ \lrb{p_{\ell}^{r_{\ell n}}} }$   
are given 
by Eq.\ \bref{2.33}.  
The dimension $d_{\ell}$ of the representation 
for the $\ell$th sector is  
\be 
 d_{\ell} = 
  \prod_{n=1}^{N_{\ell}} p_{\ell}^{r_{\ell n}}  
\quad , 
\label{2.36} 
\ee  
where 
\be  
 r_{\ell n} \equiv s_{\ell} - u_{\ell n} 
\quad .  
\label{2.37} 
\ee   
The phase factor 
in Eq.\ \bref{2.35}  
ensures that Eq.\ \bref{2.14} holds.  
It is a straightforward computation 
to verify Eq.\ \bref{2.13}.  
Hence, 
Eq.\ \bref{2.35} 
is indeed a representation of the  
$\Gamma_i^{\left( {\ell} \right)}$ matrices 
of the $\ell$th sector.  
When inserted in Eq.\ \bref{2.12}, 
a unitary representation of the group $G_d$
associated with the $d$-dimensional twisted group lattice 
is obtained
via Eqs.\ \bref{2.1} and \bref{2.2}.   
The total dimension $d_{\lrb{r}}$ of the representation is 
\be 
 d_{\lrb{r}} = \prod_{\ell=1}^{L} d_{\ell} 
\quad . 
\label{2.38} 
\ee
In the next section, 
we show that Eq.\ \bref{2.35} gives a complete set of
irreducible representations.
Hence, our solution for $\Gamma_i^{\left( {\ell} \right)}$
is unique up to equivalence.

A different representation 
is obtained for different values of the $k_{ij}$.  
This is not the case for the $k_i$, however,
since some sets $\{k_i\}$ 
lead to equivalent representations.  
Such $\{ { k_i } \}$ are related by a group $E$.  
Hence, the (equivalence classes of) 
irreducible representations are uniquely labelled
by $\{ { k_i } \} \left( { \bmod \ E } \right)$ 
and $\{ k_{ij} \}$.  

We define a group $E_{\ell}$, 
which is associated 
with the $\ell$th sector of the tensor product 
in Eq.\ \bref{2.12},  
by specifying its generators. 
There are $2 N_\ell$ generators, 
and they act on the $n$th factor 
in Eq.\ \bref{2.35} by conjugation.  
They are  
$$ 
 E_{P}^{\lrb{ \ell n} }: 
  {\hbox{ conjugation by }} 
 P_{\left( { p_{\ell}^{r_{ \ell n}} } \right)}  
  {\hbox{ on the }} 
   n{\hbox{th factor in the }} \ell{\hbox{th sector}}  
\quad , 
\phantom{(2.39)}
$$ 
\be  
  E_{Q^{-1}}^{\lrb{ \ell n} }: 
   {\hbox{ conjugation by }} 
  Q_{\left( { p_{\ell}^{r_{ \ell n}} } \right)}^{-1}  
   {\hbox{ on the }} 
   n{\hbox{th factor in the }} \ell{\hbox{th sector}}  
\quad , 
\label{2.39} 
\ee   
where $n = 1, 2, \ldots , N_{\ell}$.  
A simple calculation shows that 
the effect 
of these conjugations is to 
shift the momenta $k_i$ in the following way:    
$$ 
  {\hbox {for}}\ E_{P}^{\lrb{ \ell n} }: 
\qquad
  k_i \ \to \ k_i + 
   {{a_i^{{\lrb{ \ell n} }}  L_i} \over 
       { p_{\ell}^{r_{ \ell n}} }} 
   \pmod {L_i} 
\quad , 
$$ 
\be  
  {\hbox {for}}\ E_{Q^{-1}}^{\lrb{ \ell n} }: 
\qquad
  k_i \ \to \ k_i + 
   {{b_i^{{\lrb{ \ell n} }}  L_i} \over 
       { p_{\ell}^{r_{ \ell n}} }} 
   \pmod {L_i}  
\quad . 
\label{2.40} 
\ee    
Sets of momenta $\{ k_i \}$ which are related 
by repeated shifts of Eq.\ \bref{2.40} 
lead to equivalent representations. 
Because the 
$\vec a^{\lrb{ \ell n} }$ and 
$\vec b^{\lrb{ \ell n} }$ 
are independent,  
the number of identifications made under $E_{\ell}$
in Eq.\ \bref{2.40} is 
$ 
\prod_{n=1}^{N_{\ell}} p_{\ell}^{2 r_{\ell n}} 
     = d_{\ell}^2 
$.  

The full group $E$ is the tensor product of the groups $E_{\ell}$. 
Since different primes are relatively prime, 
the identifications 
in Eq. \bref{2.40} 
for different $\ell$ 
are independent.  
The total number of identifications under $E$ is 
\be 
 \prod_{\ell =1}^L d_{\ell}^2 = d_{\lrb{r}}^2 
\quad . 
\label{2.41} 
\ee    

\secttit {III. Proof of Irreducibility and Completeness} 
\setcounter{section}{3}   
\setcounter{equation}{0}   
\label{s:pic} 

%\hspace{\parindent} 
To prove irreducibility and completeness of the representations, 
it suffices\ctup{hamermesh}  
to verify two equations.  
The first is   
\be  
  \sum\limits_{\left( r \right)} {d_{\lrb{r}}^2}\ =\ 
    o \left( G_d \right) 
\quad ,  
\label{3.1} 
\ee         
and the second is the orthogonality of characters 
\be  
 \sum\limits_{g\in G_d} 
 {\chi^{\left( r \right)} \left( g \right)\ 
  \chi^{\left( s \right)}\left( {g^{-1}} \right)}\ =\ 
  \delta^{\left( r \right) \left( s \right)}\ 
  o\left( G_d \right) 
\quad .   
\label{3.2} 
\ee   
Recall that the representations 
are characterized by the $\{ k_{ij} \}$,   
which range from $0$ to $N_{ij}{-}1$, 
and the $\{ k_{i} \}$ modulo $E$. 

To prove Eq.\ \bref{3.1},  
note that the number of identifications 
under $E$ is $d_{\lrb{r}}^2$, 
according to 
Eq.\ \bref{2.41}.  
The dimension of a representation 
depends only on the $\{ k_{ij} \}$ 
and not on the $\{ k_{i} \}$.  
Since each $k_i$ ranges from $0$ to $L_i{-}1$, 
there are 
$$ 
 { {\prod_{i=1}^d L_i} \over {d_{\lrb{r}}^2} }  
$$ 
independent values of the $\{ k_{i} \}$.  
Consequently,  
\be 
    \sum\limits_{\lrb{ r } } {d_{\lrb{r}}^2} =  
    \sum\limits_{ \{ k_{ij} \} }
    \sum\limits_{ \left\{ {k_i} \right\} ( \bmod {E} ) } \!\!
     {d_{\lrb{r}}^2} =
  \sum\limits_{ \{ k_{ij} \} }
  {{ \prod_{i=1}^d L_i } \over {d_{\lrb{r}}^2}}\ d_{\lrb{r}}^2 = 
  \biggl( \prod_{i<j} N_{ij} \biggr)  \biggl( \prod_{i } L_i \biggr) 
   = o \left( G_d \right) 
\quad .   
\label{3.3} 
\ee  
This is Eq.\bref{3.1}. 

It is more tedious to test Eq.\ \bref{3.2}.
To distinguish the two representations 
in Eq.\ \bref{3.2}, 
we use the superscripts $(r)$ and $(s)$.  
Hence, 
the representations $\lrb{r}$ and $\lrb{s}$ 
associated with the characters  
correspond to the sets 
$  \left\{ { \{ { k_{ij}^{\left( r \right)} } \} ,
             \{ { k_{i}^{ \left( r \right)} } \}
          } \right\} 
$ 
and 
$  \left\{ { \{ { k_{ij}^{\left( s \right)} } \} ,
             \{ { k_{i}^{ \left( s \right)} } \}
          } \right\} 
$.  
We also append superscripts $(r)$ and $(s)$ 
to the other quantities obtained 
in the constructions 
of Sect.\ 2. 

Express the element $g$ 
as in Eq.\ \bref{a}. 
The sum over $g$ 
in Eq.\ \bref{3.2} 
then becomes a sum over the powers $n_{ij}$ and $n_i$: 
$$  
 \sum\limits_{g\in G_d}  
  \chi^{\lrb{ r}} \lrb{ g     }\  
  \chi^{\lrb{ s}} \lrb{ g^{-1}} 
  \qquad\qquad\qquad\qquad\qquad\qquad\qquad\qquad
  \qquad\qquad\qquad\qquad\qquad\qquad\phantom{.} 
$$ 
\be 
 =\
 \sum_{ \lrcb{n_{i}}} \sum_{\lrcb{n_{ij}}} 
  \chi^{\lrb{r}} 
  \Bigl(\gelement\Bigr)\ 
  \chi^{\lrb{s}} 
  \Bigl(\gelementinv\Bigr)
\quad .  
\label{3.4} 
\ee  
Equation \bref{2.1}  
implies  
\be  
  \chi^{\left( r \right)}\left( g \right)\ 
  \chi^{\left( s \right)}\left( {g^{-1}} \right) 
   \ \propto\  
     \exp \biggl[ {
        2\pi i\sum\limits_{i<j} 
          {{{\Delta k_{ij}\ n_{ij}} \over {N_{ij}}}}
                  } \biggr] 
\quad ,  
\label{3.5} 
\ee        
where 
$ 
  \Delta k_{ij} \equiv 
     k_{ij}^{\lrb{ r }} - k_{ij}^{\lrb{ s }}  
$.  
By summing over the $n_{ij}$ 
in Eq.\ \bref{3.4},  
one concludes that   
\be  
  \sum\limits_{\lrcb{n_{ij}}} 
   {\chi^{\left( r \right)}\left( g \right)\ 
    \chi^{\left( s \right)}\left( {g^{-1}} \right)}\ =\ 0 
\ , \quad \quad 
    \hbox{ unless } \Delta k_{ij} = 0 
    \hbox{ for all } i<j
\quad .  
\label{3.6} 
\ee   
If $ k_{ij}^{ \lrb{ r } } \ne k_{ij}^{ \lrb{ s } } $ for some $ij$, 
then the representations $\lrb{r}$ and $\lrb{s}$ are different, 
Eq.\ \bref{3.4} 
is zero, and 
Eq.\ \bref{3.2} 
is satisfied.     
Hence, for the remainder of this section  
we may assume that 
\be  
  k_{ij}^{ \lrb{ r } } = k_{ij}^{ \lrb{ s } } 
\ , \quad \quad 
    \hbox{ for all } i<j 
\quad .  
\label{3.7} 
\ee        
With all $\Delta k_{ij} = 0$, 
the sum over $n_{ij}$ 
in Eq.\ \bref{3.4} 
produces a factor of 
\be  
  \sum\limits_{\lrcb{n_{ij}}} 1\ =\
    \prod_{i<j} {N_{ij}}
\quad .  
\label{3.8} 
\ee          

Since the $k_{ij}$ are now
the same for both representations, 
almost all the quantities defined 
in Sect.\ 2 
are the same for $\lrb{r}$ and $\lrb{s}$:  
$$ 
  k_{ij}^{\prime \lrb{ r } } = 
  k_{ij}^{\prime \lrb{ s } } 
\ , \quad \quad 
  N_{ij}^{\prime \lrb{ r } } = 
  N_{ij}^{\prime \lrb{ s } } 
\ , \quad \quad
    \hbox{ for } 1 \le i<j \le d   
\quad ,  
$$  
\be  
 \Gamma_j^{ \lrb{ \ell } \lrb{ r } } = 
 \Gamma_j^{ \lrb{ \ell } \lrb{ s } }
\ , \quad \quad 
 \Omega^{ \lrb{ \ell } \lrb{ r }} = 
 \Omega^{ \lrb{ \ell } \lrb{ s }} 
\ , \quad \quad
    \hbox{ for each } \ell \hbox{ sector }   
\quad .  
\label{3.9} 
\ee          
As a consequence, 
the exponents  
in Eqs.\ \bref{2.29} and \bref{2.37} 
as well as the vector components 
in Eq.\ \bref{2.33} 
in each $\ell$ sector 
are the same:  
$$ 
 u_{\ell n}^{ \lrb{ r } } = u_{\ell n}^{ \lrb{ s } }
\ , \quad \quad 
 r_{\ell n}^{ \lrb{ r } } = r_{\ell n}^{ \lrb{ s } }
\quad ,  
$$ 
\be  
  a_i^{ \lrb{\ell n} \lrb{ r } } = 
  a_i^{ \lrb{\ell n} \lrb{ s } }  
\ , \quad  
  b_i^{ \lrb{\ell n} \lrb{ r } } = 
  b_i^{ \lrb{\ell n} \lrb{ s } }  
\ , \quad \quad
    \hbox{ for } i=1,2, \dots , d   
\quad ,  
\label{3.10} 
\ee   
for all $n$. 
The only difference between the matrices 
of the two representations 
is in the $k_j$ 
in the exponent  
of Eq.\ \bref{2.2}.  

The matrices 
$P_{(N)}$ and $Q_{(N)}$ 
in Eq.\ \bref{c} 
satisfy the property that 
$\Tr \lrb{  P_{(N)}^n Q_{(N)}^{n^{\prime}} }$ vanishes
unless 
$          n = 0 \ ( \bmod \ {N} )$ 
and    
$ n^{\prime} = 0 \ ( \bmod \ {N} )$.  
Likewise in a tensor product, 
$ 
 \Tr \lrb{ \bigotimes_m   
       Q_{ \lrb{N_m} }^{n_m} 
       P_{ \lrb{N_m} }^{n^{\prime}_m}  
  }
$   
is zero unless 
$           n_m = 0 \ ( \bmod \ {N_m} )$ 
and 
$  n^{\prime}_m = 0 \ ( \bmod \ {N_m} )$  
for all $m$.  
Using Eqs.\ \bref{2.2}, \bref{2.12} and \bref{2.35}, 
one concludes that 
the nonzero terms in the $n_i$ sums 
in Eq.\ \bref{3.4} 
occur exactly when
\be 
  \sum\limits_{i=1}^d {a_i^{\lrb{\ell n}} n_i}\ =\  
  \sum\limits_{i=1}^d {b_i^{\lrb{\ell n}} n_i}\ =\ 0 
    \pmod{ p_{\ell}^{r_{\ell n}} } 
\quad , 
\label{3.11} 
\ee        
for all $n$ and all $\ell$. 
These constraints 
need to be imposed on the $n_i$ 
in the sums 
of Eq.\ \bref{3.4}.  
When they are satisfied, one has 
\be  
  \chi^{\left( r \right)}\left( g \right)\ 
  \chi^{\left( s \right)}\left( {g^{-1}} \right)\ =\  
    d_{\lrb{r}}^2\ 
    \exp \biggl[ { 
       2\pi i\sum\limits_{i=1}^d 
       { { {\Delta k_i\ n_i} \over {L_i} } } 
                 } \biggr] 
\quad ,  
\label{3.12} 
\ee         
where 
$ 
  \Delta k_{i} \equiv 
    k_{i}^{\left( r \right)} - k_{i}^{\left( s \right)} 
$. 
Incorporating Eq.\ \bref{3.8}, we have  
\be  
  \sum\limits_{g\in G_d} 
    {\chi^{\left( r \right)}\left( g \right)\ 
     \chi^{\left( s \right)}\left( {g^{-1}} \right)}\ =\  
  d_{\lrb{r}}^2\ 
  \delta_{k_{ij}^{\left( r \right)}, k_{ij}^{\left( s \right)}}\ 
  \Bigl( \prod_{i<j} {N_{ij}} \Bigr)\ 
  \mathop{{\sum}'}_{\lrcb{n_i}} 
    \exp \biggl[  
        2\pi i\sum\limits_{i=1}^d    
      {{\Delta k_i\ n_i} \over {L_i}}  
         \biggr] 
\ ,
\label{3.13} 
\ee         
where the prime indicates that the $n_i$ sums are constrained 
by Eq.\ \bref{3.11}.  

In Ref.\ \ct{ls94a}, 
a trick was found to handle the constraints 
in Eq.\ \bref{3.11}.  
By introducing additional summation variables 
$m^{\lrb{\ell n}}$ and $\overline m^{\lrb{\ell n}}$, 
the constraints can be implemented 
by inserting into free $n_i$ sums the factor  
$$   
  \prod_{\ell} \left\{ \prod_{n=1}^{N_{\ell}}   
  \Biggl( { 
    {1 \over {p_{\ell}^{r_{\ell n}}}} 
  \sum\limits_{m^{ \lrb {\ell n} }=1 
                 }^{p_{\ell}^{r_{\ell n}}} 
     \exp \biggl[ { 
     -2\pi i{{m^{ \lrb {\ell n} }} \over 
        {p_{\ell}^{r_{\ell n}}}}\sum\limits_{i=1}^d 
     {a_{i}^{ \lrb {\ell n } }n_i} 
                 } \biggr] 
          } \Biggr)
       \right\} 
     \; \times 
$$ 
\be  
  \prod_{\ell} \left\{ \prod_{n=1}^{N_{\ell}}   
  \Biggl( { 
    {1 \over {p_{\ell}^{r_{\ell n}}}} 
  \sum\limits_{\overline m^{\lrb{ \ell n} }=1 
                }^{p_{\ell}^{r_{\ell n}}} 
     \exp \biggl[ { 
     -2 \pi i{{\overline m^{\lrb{ \ell n} }} \over 
        {p_{\ell}^{r_{\ell n}}}}\sum\limits_{i=1}^d 
     {b_{i}^{ \lrb {\ell n } }n_i} 
                 } \biggr] 
          } \Biggr)
       \right\}
\quad . 
\label{3.14} 
\ee        
Hence, we insert 
Eq.\ \bref{3.14} in Eq.\ \bref{3.13} 
and sum over the $n_i$ freely.  
When the $n_i$ sums are done before 
the $m^{\lrb{\ell n} }$ and 
$\overline m^{\lrb{\ell n} }$ sums% 
{\footnote{The interchange 
of the order of sums 
is permitted 
because all sums involve a finite number of terms.}}, 
one finds a non-zero result if and only if  
\be  
  \Delta k_i\ =\ 
 L_i \sum\limits_{\ell} \lrb{
   \sum\limits_{n=1}^{N_{\ell}} 
 \biggl( 
    {{{ m^{\lrb{\ell n} } 
    a_{i}^{\lrb{\ell n} } 
      } \over {p_{\ell}^{r_{\ell n}}}}} +  
    {{{\overline m^{\lrb{\ell n} } 
    b_{i}^{\lrb{\ell n} } 
      } \over {p_{\ell}^{r_{\ell n}}}}}  
    \biggr)  } 
      \pmod{ L_i } 
%\ , \quad 
%\hbox{ for } i = 1, 2, \dots , d 
\quad ,   
\label{3.15} 
\ee   
for $i = 1, 2, \dots , d$.       
Recall that different sectors $\ell$ 
are associated with different primes, 
which are, of course, relatively prime.   
Combining this fact  
with an 
explicit examination of 
$a_{i}^{\lrb{\ell n} }$ and 
$b_{i}^{\lrb{\ell n} }$ 
of Sect.\ 2 
reveals that there is at most one  
$ m^{\lrb{\ell n} }$ 
and one 
$ \overline m^{\lrb{\ell n} } $ 
which solve Eq.\ \bref{3.15}.  
Hence, at best, one term 
among the auxiliary summation variables 
$ m^{\lrb{\ell n} }$
and  
$ \overline m^{\lrb{\ell n} } $
contributes.  
On the other hand, 
if the momenta 
$ k_{i}^{\left( r \right)}$ and 
$ k_{i}^{\left( s \right)}$ 
differ as in  
Eq.\ \bref{3.15}, 
then they are related by an element of $E$ 
involving the generators raised to 
$ m^{\lrb{\ell n} }$th
and  
$ \overline m^{\lrb{\ell n} } $th
powers, 
\be  
  \prod_{\ell} \Biggl\{ \prod_{n=1}^{N_{\ell}}   
  \left( 
     \left[   
         E_{P}^{ \lrb{\ell n} } 
     \right]^{m^{ \lrb {\ell n}}}  
     \left[   
         E_{Q^{-1}}^{ \lrb{\ell n} }  
     \right]^{\overline m^{ \lrb {\ell n}}} 
  \right)
\Biggr\}
\quad ,   
\label{3.16} 
\ee        
as can be seen 
from Eqs.\ \bref{2.39} and \bref{2.40}.  
We have thus shown that 
$  
  (r) \sim  (s)  
$ 
if a non-zero result 
in Eq.\ \bref{3.2} 
is to be obtained.  

It remains to check whether 
the normalization 
in Eq.\ \bref{3.2} 
is correct when 
$\lrb{r}$ is equivalent to $\lrb{s}$. 
When $\lrb{r} \sim \lrb{s}$,  
there is a unique non-zero term 
in the 
$ m^{ \lrb{\ell n} }$ 
and  
$ \overline m^{ \lrb{\ell n} }$ 
sums which contributes.  
It satisfies 
Eq.\ \bref{3.15}.   
For this term, 
the phase factors 
in Eqs.\ \bref{3.13} and \bref{3.14} 
cancel.  
Then the sums over the ${n_i}$ 
merely produce 
$$
 \prod_{i=1}^{d} {L_i} 
\quad , 
$$ 
and the $1/p_{\ell}^{r_{\ell n}}$ factors 
in Eq.\ \bref{3.14} 
give   
$$ 
  \prod_\ell 
  \fr{1}{ \prod_{n=1}^{N_{\ell}} 
                 p_{\ell}^{2 r_{\ell n}} }\ =\ 
  \fr{1}{d_{\lrb{r}}^2}  
\quad . 
$$   
When these results 
are inserted into 
Eq.\ \bref{3.13},  
one finds   
$$ 
  \sum\limits_{g\in G_d} 
   {\chi^{\left( r \right)}\left( g \right)\ 
    \chi^{\left( s \right)}\left( {g^{-1}} \right)}\ =\ 
     d_{\lrb{r}}^2\ \delta^{\left( r \right)\left( s \right)}\ 
      {{\prod_{i=1}^{d} {L_i}} \over {d_{\lrb{r}}^2}}\ 
    \Bigl( {\prod_{i<j} {N_{ij}}} \Bigr)
    \qquad\qquad\qquad\qquad\qquad\phantom{.}
$$ 
\be  
    \phantom{.}\qquad\qquad\qquad 
      =\ \delta^{\left( r \right)\left( s \right)}\
    \Bigl( { \prod_{i=1}^{d} {L_i} }\Bigr) 
    \Bigl( {\prod_{i<j} {N_{ij}}} \Bigr)\ =\  
    \delta^{\left( r \right) \left( s \right)}\ 
    o\left( G_d \right) 
\quad .  
\label{3.17} 
\ee         
This is Eq.\ \bref{3.2}.

\secttit{IV. Conclusion} 
\setcounter{section}{4}   
\setcounter{equation}{0}   
\label{s:c} 

In this work, 
we have obtained the representations of the group 
associated with the general $d$-dimensional twisted group lattice. 
A proof that   
all the irreducible representations have been found 
was given in Sect.\ 3.  

Our work solves 
the general $d$-dimensional twisted group lattice.  
The partition function $Z$ 
for a free charged bosonic theory is 
given by 
Eqs.\ (4.5)--(4.7) 
of Ref.\ \ct{ls94a}, 
\ie, 
\be  
  Z\ =\ \prod\limits_{ \lrb{r} }  
  \left[ { 
    \det \biggl( { 
           \sum\limits_{h \in \NN_e}  
   \lambda_h\ D^{ \lrb{r} } \left( h \right)  
                } \biggr) 
          } \right]^{ -d_{ \lrb{r} } } 
\quad , 
\label{4.1} 
\ee       
where 
$$ 
  \sum\limits_{h\in \NN_e} 
   \lambda_h\ D^{ \lrb{r}} \left( h \right)\ =\ 
   \lambda_e\ I_{\lrb{d_{ \lrb{r}} }}  
  \qquad\qquad\qquad\qquad\qquad\qquad\qquad\qquad\qquad\qquad
$$ 
\be  
  \phantom{.}\qquad\qquad
  +\ \sum\limits_{j=1}^d  
  \left[ 
   \lambda_h\ \Gamma_j\ 
   \exp \biggl( { 2\pi i k_j 
       \over L_j } \biggr) + 
   \lambda_{h^{-1}}\ \Gamma_j^{\dagger}\ 
   \exp \biggl( { -2\pi i k_j 
       \over L_j } \biggr) 
  \right] 
\quad ,   
\label{4.2} 
\ee   
and where $\NN_e = \{ e \} \cup \NN$, 
$\lambda_e$ is a mass parameter, 
and the $\lambda_h$ for $h \in \NN$ are hopping parameters.  
Here, $\{ k_{ij} , k_i \}$ are the representation labels 
associated with $\lrb{r}$.
The gamma matrices $\Gamma_j$ in 
Eq.\ \bref{4.2}  
are given 
in Eqs.\ \bref{2.12} and \bref{2.35}.  
The product in Eq.\ \bref{4.1} is over 
all the irreducible representations $\lrb{r}$ of $G_d$, \ie, 
\be  
    \prod_{\lrb{r}}\ =\ 
    \prod_{\left\{ {k_{ij}=0} \right\}}^{\left\{ {N_{ij}-1} \right\}}  
    \ \prod_{ \{ k_i \} } \pmod {E}   
\quad ,  
\label{4.3} 
\ee         
where the group $E$ 
is specified 
in Eq.\ \bref{2.40}.  
The partition function $Z$ corresponds to 
a gas of closed oriented loops, 
where the loops can be regarded as particle trajectories.  
The loops must statisfy the constraint that 
the area projected onto the $i$-$j$ plane 
has zero area $\lrb{ \bmod \ N_{ij} }$, 
for all $i < j$.  
To perform a perturbative expansion 
of an interacting theory, 
one needs the propagator.  
It involves the inverse 
of the matrix 
in Eq.\ \bref{4.2}.  
The precise formula is given 
in Eq.\ (4.9) 
of Ref.\ \ct{samuel90a}.    
The partition function for the free fermionic case
is given by 
Eq.\ \bref{4.1} 
with the exponent $-d_{ \lrb{r} }$ 
replaced by $d_{ \lrb{r} }$.  

The solution methods obtained 
in Refs.\ \ct{samuel91b,samuel92a,ls94a} 
for the $d=2$, $d=3$ and $d=4$ cases
appear to be quite different.  
However, 
it can be verified that the two-form approach 
of the current work 
reproduces the constructions for these cases.  
Our method provides a unifying framework 
for understanding these systems for different dimensions.  

The twisted lattices can be used as discrete models 
of Euclidean space-time.\ctup{ls94a}   
When all $L_i$ are equal and all $N_{ij}$ are equal, 
the twisted lattices possess the same discrete rotational 
symmetries as regular lattices.  
This was shown in 
Ref.\ \ct{ls94a}.  
Compared to regular lattices, 
discrete translations no longer commute.  
Hence, 
a twisted group lattice corresponds 
to a non-commutative geometry  
of a discrete Euclidean space-time.   

Finally, we would like to remark that 
because Eq.\ \bref{ab} 
appears in many contexts, 
it is likely that our construction method 
for the case 
in which the dimension of the matrices $\Gamma_i$ is arbitrary  
will find future uses.  

In summary, 
we have solved 
the general $d$-dimensional twisted group lattice 
and provided a few applications.  
Further investigations will hopefully uncover 
additional uses for our research.

\secttit{Acknowledgements} 

We thank 
H.\ Skarke for discussions
and
J.\ Wess 
for his hospitality 
at the Max-Planck-Institut f\"ur Physik 
where some of this work was carried out.  
This work is supported in part  
by a NATO Collaborative Research Grant 
(grant number CRG 930761),  
by the United States Department of Energy 
(grant number DE-FG02-92ER40698), 
by the Alexander von Humboldt Foundation, 
and by the PSC Board of Higher Education at CUNY. 

\vfill\eject

\secttit{Appendix} 
\label{s:a} 
\setcounter{section}{5}   
\def\theequation{A.\arabic{equation}}
\setcounter{equation}{0}   

In this appendix, 
we illustrate the method of constructing the $\Gamma_i$  
of Sect.\ 2 
for a six-dimensional twisted lattice.  
Since the $\Gamma_i$ depend only on 
$k_{ij}^{\prime}$ and $N_{ij}^{\prime}$,   
it suffices to specify these primed quantities. 
We select  
$$ 
  k_{13}^{\prime} = 
  k_{14}^{\prime} = 
  k_{23}^{\prime} = 
  k_{24}^{\prime} = 
  k_{26}^{\prime} = 
  k_{56}^{\prime} = 0 
\quad , 
$$ 
$$ 
  k_{12}^{\prime} = 
  k_{15}^{\prime} = 
  k_{16}^{\prime} = 
  k_{25}^{\prime} = 
  k_{34}^{\prime} = 
  k_{35}^{\prime} = 
  k_{45}^{\prime} = 
  k_{46}^{\prime} = 1 
\quad , 
$$ 
$$ 
  k_{36}^{\prime} = 3 
\quad , 
$$ 
$$ 
  N_{13}^{\prime} = 
  N_{14}^{\prime} = 
  N_{23}^{\prime} = 
  N_{24}^{\prime} = 
  N_{26}^{\prime} = 
  N_{56}^{\prime} = 1 
\quad , 
$$ 
$$ 
  N_{12}^{\prime} = 
  N_{15}^{\prime} = 
  N_{16}^{\prime} = 
  N_{25}^{\prime} = 
  N_{45}^{\prime} = 
  N_{46}^{\prime} = 2 
\quad , 
$$ 
\be  
  N_{34}^{\prime} = 
  N_{35}^{\prime} = 
  N_{36}^{\prime} = 4 
\quad .  
\label{A.1} 
\ee  
The least common multiple $N'$ of the $N_{ij}^{\prime}$ 
is $4$, 
so that we are dealing with a single prime case.  
It is therefore not necessary to specify the prime sector, 
and we drop the label $\ell$.  
The quantities 
in Eq.\ \bref{2.8} are 
\be  
  N' = 2^2 
\ , \quad \quad 
   p=2 
\ , \quad \quad 
   s=2 
\quad . 
\label{A.2} 
\ee 

{}From Eqs.\ \bref{A.1} and \bref{A.2},  
one obtains the  
$ \omega_{ij} $.  
Then, the matrix $\omega=(\omega_{ij})$ is computed 
from Eq.\ \bref{2.15}.   
One finds  
\be  
  \omega\ = 
  \lrb{ 
    \matrix{ 
   0  &2  &0  &0  &2  &2  \cr
   2  &0  &0  &0  &2  &0  \cr
   0  &0  &0  &1  &1  &3  \cr
   0  &0  &3  &0  &2  &2  \cr
   2  &2  &3  &2  &0  &0  \cr
   2  &0  &1  &2  &0  &0  \cr
          } 
      } 
    \pmod {4} 
\quad . 
\label{A.3} 
\ee 

The next step is to achieve the decomposition 
of Eq.\ \bref{2.30}. 
The entry $\omega_{34}$ satisfies 
$\omega_{34} \ne 0 \lrb{\bmod \ 2}$.  
For the symplectic construction,
it was assumed that 
$\omega_{12} \ne 0 \lrb{\bmod \ 2}$.   
Hence, one performs the label interchanges 
$ 
  1 \leftrightarrow 3 
$,  
$ 
  2 \leftrightarrow 4 
$, 
constructs 
the $a_i^{\lrb{1}}$ and the $b_i^{\lrb{1}}$ 
using the method of the Lemma, 
and then interchanges the labels back. 
The results are 
$ 
  a_3^{\lrb{1}} = 1 
$,
$ 
  a_4^{\lrb{1}} = 0 
$, 
$ 
  a_1^{\lrb{1}} = -\omega_{41}/\omega_{34} = 0 
$, 
$ 
  a_2^{\lrb{1}} = -\omega_{42}/\omega_{34} = 0 
$, 
$ 
  a_5^{\lrb{1}} = -\omega_{45}/\omega_{34} = -2 = 2\lrb{\bmod \ 4} 
$ 
$, 
  a_6^{\lrb{1}} = -\omega_{46}/\omega_{34} = -2 = 2\lrb{\bmod \ 4} 
$,  
$ 
  b_3^{\lrb{1}} = 0 
$, 
$ 
  b_4^{\lrb{1}} = \omega_{34} = 1 
$, 
$ 
  b_1^{\lrb{1}} = \omega_{31} = 0 
$, 
$ 
  b_2^{\lrb{1}} = \omega_{32} = 0 
$,  
$ 
  b_5^{\lrb{1}} = \omega_{35} = 1 
$, 
$ 
  b_6^{\lrb{1}} = \omega_{36} = 3 
$. 
Hence, the one-forms 
in Eq.\ \bref{2.26} are 
\be  
  \alpha^{\lrb{1}}\ =\ dx^3 + 2dx^5 + 2dx^6 
\ , \quad \quad 
  \beta^{\lrb{1}}\ =\ dx^4 + dx^5 + 3dx^6 
\quad , 
\label{A.4} 
\ee 
and the corresponding vectors 
$\vec a^{\lrb{1}}$ and $\vec b^{\lrb{1}}$ read 
\be  
  \vec a^{\lrb{1}} = \left( {0,0,1,0,2,2} \right) 
\ , \quad \quad 
  \vec b^{\lrb{1}} = \left( {0,0,0,1,1,3} \right)
\quad . 
\label{A.5} 
\ee 
Following Eq.\ \bref{2.27}, 
$
  \hat \Omega = 
       \Omega - \alpha^{\lrb{1}}\wedge \beta^{\lrb{1}} 
$.  
A straightforward computation gives 
the corresponding matrix $\hat \omega$:  
\be  
  \hat \omega\ = 
  \lrb{ 
    \matrix{ 
   0  &2  &0  &0  &2  &2  \cr
   2  &0  &0  &0  &2  &0  \cr
   0  &0  &0  &0  &0  &0  \cr
   0  &0  &0  &0  &0  &0  \cr
   2  &2  &0  &0  &0  &0  \cr
   2  &0  &0  &0  &0  &0  \cr
          } 
      } 
   \pmod{4} 
\quad . 
\label{A.6} 
\ee 
The entries involving indices $3$ and $4$ are zero, 
as expected 
since we have performed the construction procedure 
using $\omega_{34}$.  

Note that $\hat \omega$  
is divisible by $2$.  
Hence,  
the remainder form $\hat\Omega$ equals $\Omega'_0$,
and
\be  
  \Omega\ =\ 
  \alpha^{\lrb{1}} \wedge \beta^{\lrb{1}} + 
  \Omega'_0 \pmod{4} 
\quad . 
\label{A.7} 
\ee 
Following Eq.\ \bref{2.30}, 
we construct $\Omega_1$ by dividing by $p$: 
$ 
 \Omega_1 = \Omega'_0 /2 
$. 
The corresponding matrix $\omega_1\equiv\tilde\omega$ is  
\be  
  \tilde\omega\ =  
  \lrb{ 
    \matrix{ 
   0  &1  &0  &0  &1  &1  \cr
   1  &0  &0  &0  &1  &0  \cr
   0  &0  &0  &0  &0  &0  \cr
   0  &0  &0  &0  &0  &0  \cr
   1  &1  &0  &0  &0  &0  \cr
   1  &0  &0  &0  &0  &0  \cr
          } 
      } 
     \pmod{2} 
\quad . 
\label{A.8} 
\ee 
The entry $\tilde\omega_{12}$ satisfies 
$ 
  \tilde\omega_{12} \ne 0 \lrb{\bmod \ 2}
$.  
We use this entry to construct 
$a_i^{\lrb{2}}$ and $b_i^{\lrb{2}}$.    
The method of 
Sect.\ 2 
gives 
$ 
  a_1^{\lrb{2}} = 1 
$, 
$ 
  a_2^{\lrb{2}} = 0 
$,  
$ 
  a_3^{\lrb{2}} = 
    -\tilde\omega_{23}/\tilde\omega_{12} = 0 
$,  
$ 
  a_4^{\lrb{2}} = 
    -\tilde\omega_{24}/\tilde\omega_{12} = 0 
$,  
$ 
  a_5^{\lrb{2}} = 
    -\tilde\omega_{25}/\tilde\omega_{12} = -1 = 
         1 \lrb{\bmod \ 2} 
$,  
$ 
  a_6^{\lrb{2}} = 
    -\tilde\omega_{26}/\tilde\omega_{12} = 0 
$, 
$ 
  b_1^{\lrb{2}} = 0 
$,  
$ 
  b_2^{\lrb{2}} = \tilde\omega_{12} = 1 
$,  
$ 
  b_3^{\lrb{2}} = \tilde\omega_{13} = 0 
$,  
$ 
  b_4^{\lrb{2}} = \tilde\omega_{14} = 0 
$,  
$ 
  b_5^{\lrb{2}} = \tilde\omega_{15} = 1 
$,  
$ 
  b_6^{\lrb{2}} = \tilde\omega_{16} = 1 
$.   
These results are summarized as  
\be  
  \alpha^{\lrb{2}}\ =\ dx^1 + dx^5 
\ , \quad  \quad  
  \beta^{\lrb{2}}\ =\ dx^2 + dx^5 + dx^6 
\quad ,  
\label{A.9} 
\ee 
or 
\be 
  \vec a^{\lrb{2}} = \left( {1,0,0,0,1,0} \right) 
\ , \quad  \quad  
  \vec b^{\lrb{2}} = \left( {0,1,0,0,1,1} \right)
\quad .  
\label{A.10} 
\ee 

The next remainder form, $\hat \Omega_1$, is calculated using 
$ 
  \hat \Omega_1 = 
   \Omega_1 - \alpha^{\lrb{2}}\wedge \beta^{\lrb{2}} 
$.  
One finds that the corresponding matrix 
$\hat \omega_1$ is  
\be  
  \hat \omega_1\ =   
  \lrb{ 
    \matrix{ 
   0  &0  &0  &0  &0  &0  \cr
   0  &0  &0  &0  &0  &0  \cr
   0  &0  &0  &0  &0  &0  \cr
   0  &0  &0  &0  &0  &0  \cr
   0  &0  &0  &0  &0  &1  \cr
   0  &0  &0  &0  &1  &0  \cr
          } 
      } 
     \pmod{2}  
\quad ,   
\label{A.11} 
\ee 
so that 
\be  
  \hat \Omega_1\ =\ \alpha^{\lrb{3}} \wedge \beta^{\lrb{3}} 
\quad ,  
\label{A.12} 
\ee 
where 
\be  
  \alpha^{\lrb{3}}\ =\ dx^5 
\ , \quad  \quad  
  \beta^{\lrb{3}}\ =\ dx^6 
\quad ,  
\label{A.13} 
\ee 
and 
\be  
  \vec a^{\lrb{3}} = \left( {0,0,0,0,1,0} \right) 
\ , \quad  \quad  
  \vec b^{\lrb{3}} = \left( {0,0,0,0,0,1} \right) 
\quad .   
\label{A.14} 
\ee 
The computation terminates since all two-forms have been
decomposed completely.

In summary, 
\be  
  \Omega\ =\ 
   \alpha^{\lrb{1}} \wedge \beta^{\lrb{1}} + 
   2 \left( {\alpha^{\lrb{2}} \wedge \beta^{\lrb{2}} + 
   \alpha^{\lrb{3}} \wedge \beta^{\lrb{3}}} \right) \pmod{4}  
\quad ,    
\label{A.15} 
\ee 
where the $\alpha^{\lrb{n}}$ and the $\beta^{\lrb{n}}$ 
are given 
in Eqs.\ \bref{A.4}, \bref{A.9} and \bref{A.13}.  
{}From Eq.\ \bref{A.15}, 
one concludes that 
$M_0=1$ and $M_1=2$,
implying 
$u_1=0$ and $u_2=u_3=1$.
Hence,
the powers $r_n=2-u_n$ are 
$r_1 = 2$, $r_2 = 1$, and $r_3 = 1$,  
and the sizes of the matrices 
in the tensor products are 
$2^2 = 4$, $2^1 = 2$ and $2^1 = 2$.  
The dimension of the representation is $16$.  
Equation \bref{A.15} 
is the result of the decomposition \bref{2.30} 
for the two-form $\Omega$ associated with $\omega$ 
in Eq.\ \bref{A.3}.  

The matrices $\Gamma_i$ are constructed using 
Eq.\ \bref{2.35}:   
\begin{eqnarray}
%\be
 \matrix{
  \Gamma_1\ =\ I_{\lrb{4}} \otimes Q_{\lrb{2}} \otimes I_{\lrb{2}} 
\quad , \quad\hfill &
  \Gamma_2\ =\ I_{\lrb{4}} \otimes P_{\lrb{2}} \otimes I_{\lrb{2}} 
\quad , \quad\hfill \cr
  & \cr
  \Gamma_3\ =\ Q_{\lrb{4}} \otimes I_{\lrb{2}} \otimes I_{\lrb{2}} 
\quad , \quad\hfill &
  \Gamma_4\ =\ P_{\lrb{4}} \otimes I_{\lrb{2}} \otimes I_{\lrb{2}} 
\quad , \quad\hfill \cr
  & \cr
  \Gamma_5\ =\ Q_{\lrb{4}}^2P_{\lrb{4}} 
                 \otimes Q_{\lrb{2}}P_{\lrb{2}} \otimes Q_{\lrb{2}} 
\quad , \quad\hfill &
  \Gamma_6\ =\ Q_{\lrb{4}}^2P_{\lrb{4}}^3 
                 \otimes P_{\lrb{2}} \otimes P_{\lrb{2}} 
\quad . \quad\hfill \cr }
\nonumber \\
%\quad
\label{A.16} 
%\ee
\end{eqnarray}
The powers of the $Q$ and $P$ matrices 
for the $n$th tensor factor 
are the components of the vectors 
$\vec a^{\lrb{n}}$ 
and 
$\vec b^{\lrb{n}}$, 
which were given 
in Eqs.\ \bref{A.5}, \bref{A.10} and \bref{A.14}. 
The phases 
in Eq.\ \bref{2.35} 
for $\Gamma_5$ and $\Gamma_6$ 
in Eq.\ \bref{A.16} 
can be dropped 
for this example 
because $4$ must divide $L_5$ and $L_6$.   
In other words, 
$\Gamma_i^{L_i} = I_{\lrb{16}}$ holds 
for the $\Gamma_i$ 
in Eq.\bref{A.16} 
with or without the phase factor 
in Eq.\ \bref{2.35}.  
It is a useful exercise to verify that 
the $\Gamma_i$ do indeed satisfy 
$$ 
  \Gamma_j\ \Gamma_i\ =\ 
  \Gamma_i\ \Gamma_j\ 
   \exp \left( { 2 \pi i{{k_{ij}^{\prime}} 
       \over {N_{ij}^{\prime}}}} \right) 
  \qquad\hbox{ for }j>i 
\quad , 
$$ 
with the $k_{ij}^{\prime}$ and $N_{ij}^{\prime}$ chosen
in Eq.\ \bref{A.1}.  

%\vfill\eject

\end{document}